\documentclass[final,numberedheadings]{aipproc}
\layoutstyle{8x11single}
\usepackage{amssymb,stmaryrd}
\def\op#1{\hat{#1}}
\def\vec#1{\mathbf{#1}}
\def\RR{\mathbb{R}}
\def\ket#1{|#1\rangle}
\def\norm#1{|| #1 ||}
\def\ave#1{\langle #1\rangle}
\newif\ifpdflatex\pdflatextrue
\makeatletter\@ifundefined{pdfoutput}{\pdflatexfalse}\makeatother
\def\myincludegraphics[#1]#2#3{%
\ifpdflatex \includegraphics[#1]{#2}
\else       \includegraphics[#1]{#3}
\fi}

\begin{document}
\title{Experimental Hamiltonian Identification for Qubits subject to
       Multiple Independent Control Mechanisms.}
\author{Sonia G.~Schirmer}{
  address={Dept of Applied Maths and Theoretical Physics, 
           University of Cambridge, \\ Wilberforce Rd, 
	   Cambridge, CB3 0WA, United Kingdom}}
\author{Avinash Kolli}{
  address={Dept of Applied Maths and Theoretical Physics, 
           University of Cambridge, \\ Wilberforce Rd, 
	   Cambridge, CB3 0WA, United Kingdom}}
\author{Daniel K.~L.~Oi}{
  address={Dept of Applied Maths and Theoretical Physics, 
           University of Cambridge, \\ Wilberforce Rd, 
	   Cambridge, CB3 0WA, United Kingdom}}
\author{Jared H.~Cole}{
  address={Centre for Quantum Computer Technology, School of Physics,
           University of Melbourne, Melbourne, Australia}}
\begin{abstract}
We consider a qubit subject to various independent control mechanisms and 
present a general strategy to identify both the internal Hamiltonian and
the interaction Hamiltonian for each control mechanism, relying only on  
a single, fixed readout process such as $\op{\sigma}_z$ measurements.
\end{abstract}

\maketitle


\section{Introduction}

Realizing the ultimate goal of quantum information processing, namely building
a working quantum computer, is to a large extent a problem of finding ways to
control the dynamics of a quantum system precisely.  A crucial prerequisite for
this task is one's ability to accurately determine of the dynamics of the 
physical system and its response to external (control) fields.  Quantum process
tomography (QPT), by providing a general procedure to identify the unitary (or 
completely positive) processes acting on a system, addresses this problem but 
does not solve it completely.

One problematic aspect of QPT is the assumption that one can experimentally 
determine the expectation values of a complete set of observables, or at least
perform \emph{arbitrary} single qubit measurements on a register of $n$ qubits.  
Most QIP proposals rely on a single readout process, i.e., measurement in a 
fixed basis.  For example, qubits encoded in internal electronic states of 
trapped ions or neutral atoms are read out by quantum jump detection via a 
cycling transition; readout for solid-state qubits based on Cooper-pair boxes, 
Josephson junctions, or electrons in double-well potentials usually involves
charge localization using single electron transistors or similar devices.  
Finally, solid state architectures based on electron or nuclear spin qubits
are expected to be limited to $\sigma_z$ measurements via spin-charge transfer.

It is usually assumed that local projective measurements in a fixed basis are 
sufficient since arbitrary single qubit measurements can then be realized by 
performing a local unitary transformation before measuring to achieve a change
of basis.  However, implementing such a basis change requires precise knowledge 
of the dynamics of each individual qubit and its response to control fields in
the first place, the very information we seek to determine experimentally, and 
which may be difficult to \emph{predict} precisely based on theoretical models
and computer simulations alone, in particular for systems that are sensitive 
to fabrication variance.  A possible solution to this seemingly intractable 
problem was described in~\cite{PRA69n050306}.  In the following we outline the
general strategy, discuss various ways of extracting the system parameters from
noisy experimental data, and illustrate the key steps using examples with 
simulated measurement data.

\section{General Strategy for Hamiltonian Identification}

\begin{figure}
\myincludegraphics[height=2in]{figures/pdf/Fig2.pdf}{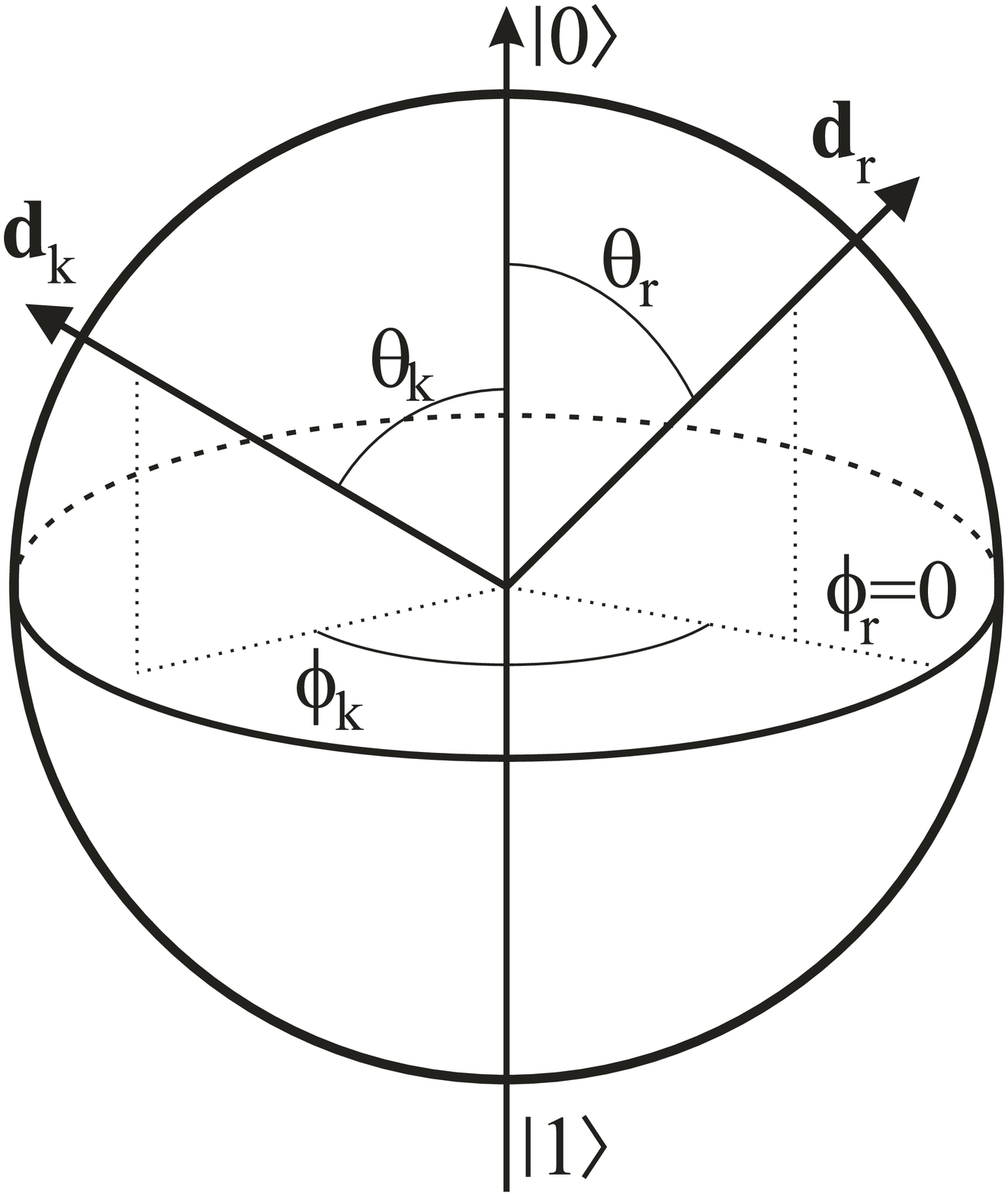} 
\hspace{1in}
\myincludegraphics[height=2in]{figures/pdf/Fig1.pdf}{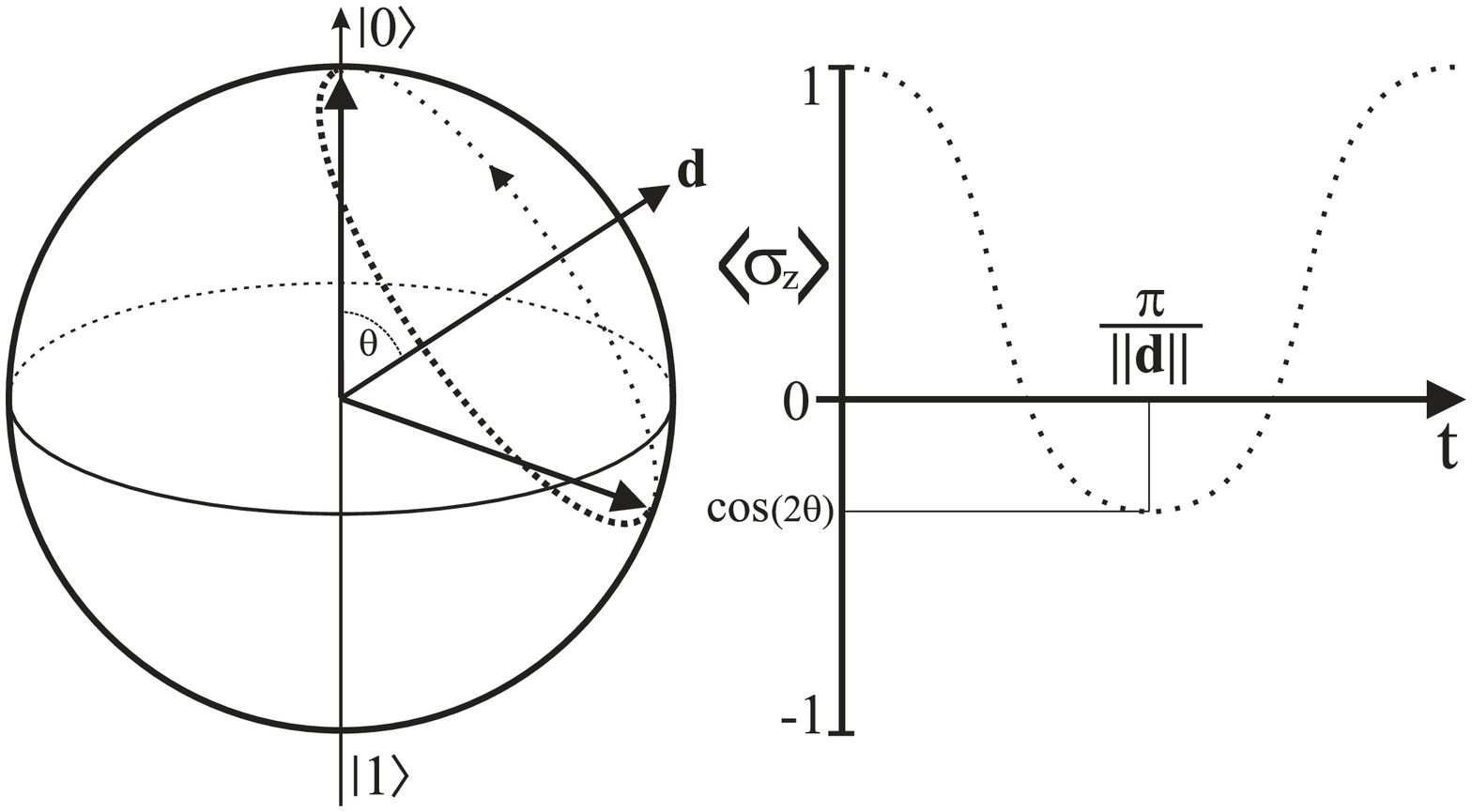}  
\caption{Left: Declinations $\theta$ and relative azimuthal angles $\phi$ for
two rotation axes.  Right: Determination of the rotation frequency and angle 
$\theta$ of the rotation axis by mapping the precession of the state $\ket{0}$.} \label{fig1}
\end{figure}

The state of a two-level system can be mapped to a Bloch vector $\vec{s}$, i.e., 
a real vector in $\RR^3$, with pure states corresponding to points on the Bloch
sphere, i.e., the surface of the unit ball.  On timescales sufficiently short 
compared to the decoherence time, the evolution of the system is governed by a
Hamiltonian, which can be written in terms of the Pauli matrices $\op{\sigma}_*$
for $*\in\{x,y,z\}$, $2\op{H}=d_0\op{I}+d_x\op{\sigma}_x+d_y\op{\sigma}_y+d_z
\op{\sigma}_z$, where $d_0$, $d_x$, $d_y$ and $d_z$ are real constants.  If the
Hamiltonian remains constant for $t_0\le t\le t_1$, the Bloch vector undergoes 
a rotation about the axis $\vec{d}=(d_x,d_y,d_z)$.  The length $\norm{\vec{d}}$
of this vector determines rotation frequency; the rotation axis can be specified
by a unit vector $\op{\vec{d}}=(\sin\theta\cos\phi,\sin\theta\sin\phi,\cos\theta)^T$, 
i.e., by two angles $\theta$ and $\phi$ as shown in Fig.~\ref{fig1}.  To identify
the parameters $d_x$, $d_y$ and $d_z$ of the Hamiltonian it therefore suffices to 
determine the rotation frequency and the angles $\theta$ and $\phi$.  Since $d_0$
results only in an unobservable global phase factor, it can be ignored.  

If the system can be repeatedly initialized in a known state, e.g., one of the
measurement basis states $\ket{0}$ or $\ket{1}$, and then measured after having
evolved for progressively longer time periods, we can map the trajectory of the 
$z$-component of the Bloch vector, and extract the frequency and angle $\theta$
of the rotation as shown in Fig.~\ref{fig1} (right).  For some systems such as 
NMR-based schemes this may be sufficient as the phase relationship between $d_x$
and $d_y$ is fixed by the phase relationship between the control fields.  In 
general, however, a second series of measurements is necessary to determine the 
horizontal angle $\phi$ of the rotation axis with respect to a reference axis
$\vec{d}_r$.  This procedure is given in Ref.~\cite{PRA69n050306} and involves 
using the values of $\theta$ and $\norm{\vec{d}}$ determined in the first step 
to select a different initial state, and mapping its precession about the desired 
rotation axis.

For a system subject to various control fields $f_m$ (e.g., fields associated with
different control electrodes) in addition to its free evolution, we must determine
both its internal Hamiltonian $\op{H}_0$ and the interaction Hamiltonian $\op{H}_m$
for each independent control mechanism.  Although we usually cannot determine the 
interaction Hamiltonians directly since we cannot switch off the internal dynamics, 
we can identify the rotation axis $\vec{d}_{0+m}^{(k)}=\vec{d}_0+f_m^{(k)}\vec{d}_m$ 
corresponding to the evolution of the system under the Hamiltonian $\op{H}_0+f_m^{(k)}
\op{H}_m$ for a fixed control setting $f_m=f_m^{(k)}$.  Repeating this procedure for
each available control field $f_m$ with several control field settings $f_m^{(k)}$ 
then allows us to extract both the internal and interaction parts of the Hamiltonian
provided that the dependence of the Hamiltonian on the control fields is linear.
(Nonlinear field effects require additional correction terms, and we will exclude 
this case in this paper.)

\section{Mapping \lowercase{$z(t)$} and Extracting the Relevant Data}
\label{sec:map}

A crucial factor in the Hamiltonian identification strategy outlined above is the
mapping of the evolution of $z(t)$ as the system precesses around a fixed rotation
axis.  The accuracy with which we can identify the relevant parameters such as the
rotation frequency and angle $\theta$ depends on the total length of time $t_f$ for
which $z(t)$ is mapped, the time resolution $\Delta t$ and the uncertainty of each
data point $z(t_k)=\ave{\sigma_z(t_k)}$, which depends on the number of times $N_e$
each experiment is repeated to obtain the ensemble average (in addition to the 
frequency of measurement errors etc).  The total number of measurements required 
to map $z(t)$ is thus $N_T=N_et_f/\Delta t$.  The choice of $\Delta t$, $t_f$ and 
$N_e$ will depend both on the system to be characterized and the method of data 
extraction to be used.

One possible approach is to use a small time step $\Delta t$ and a large number
of repetitions $N_e$ to obtain a dense set of accurate data points for a period
of time covering at least one quarter of the rotation period, and fit a cosine 
segment to the data.  Since the time we have to monitor the evolution of $z(t)$ 
is limited by the decoherence time of the system, this approach may be useful 
for systems that decohere rapidly because it requires only sampling over a short 
period of time.

Another approach, explained in detail in~\cite{PRA69n050306}, is to use relatively
coarse time sampling, with a moderate number of repetitions for each data point, 
over several rotation periods to obtain a rough estimate of the rotation frequency
using the discrete Fourier transform, followed by a second step of acquiring more
accurate additional data points in the region where the first minimum of $z(t)$ is
expected based on the first estimate, and fitting a parabola to the new data to 
find the rotation frequency and declination of the rotation axis.

A third alternative~\cite{jared} is to eliminate curve-fitting altogether and extract 
all the required information directly from the Fourier transform.  This method is 
rather elegant and does not require the high time resolution $\Delta t$ and measurement
repetition rates $N_e$ usually necessary for curve-fitting methods.  All we need
to avoid aliasing is that $\Delta t$ be less than half the rotation period $T$. 
Ensuring that this condition is satisfied requires a rough a priori estimate of $T$ 
but this should normally not be a problem.  It also permits easy estimation of the 
accuracy of the parameters.  However, since the frequency resolution $\Delta\omega=
2\pi/t_f$, we must be able to map the data for a least two complete cycles to be 
able to extract the rotation frequency, and more cycles will be required to obtain
a clearly defined peak in the frequency spectrum.  Hence, this method will be most
suited to systems whose decoherence time is sufficiently long to allow mapping of 
the evolution of $z(t)$ over several cycles.

\section{Illustrative Example}

To demonstrate the procedure, we choose a test system with $\vec{d}_0=(0.2,0,0.1)^T$, 
$\vec{d}_1=(1,1,0)^T$ and $\vec{d}_2=(0,0,1)^T$.  Fig.~2 illustrates how 
we identify the rotation frequencies and declination angles for $f_1=0$ and $f_2=
0.1$ following the 3rd approach outlined in Sec.~\ref{sec:map}.  We sample $z(t)$ 
over several rotation periods with an intermediate time step ($\Delta t=0.25$) and 
a small number of repetitions ($N_e=10$) for each measurement.  We then obtain an
estimate of the rotation frequency by taking the Fourier transform of the data and 
finding the frequency $\omega_p$ such that $|F(\omega_p)|=\max_{n>0} |F(\omega_n)|$, 
where $F(\omega_n)$ is the $n$th Fourier coefficient.  Since the sampling period is
usually not an integer multiple of the rotation period, the peak in the Fourier 
spectrum will tend to be unsharp, and our estimate inaccurate.  To improve it, we 
compute the function $P(t_f)=[F_{t_f}(\omega_p)-F_{t_f}(\omega_{p-1})-F_{t_f}
(\omega_{p+1})]/[F_{t_f}(\omega_{p-1})+F_{t_f}(\omega_{p+1})]$ where $F_{t_f}$ is 
the Fourier transform of the truncated data for $0\le t \le t_f$ and $\omega_p$ is 
the frequency where the first peak in the spectrum (excl.\ $F(0)$) occurs, $|F_{t_f}
(\omega_p)|=\max_{n>0} |F_{t_f}(\omega_n)|$.  $P(t_f)$ assumes a maximum when $t_f$ 
is an integer multiple of the rotation period, thus allowing us to find the optimal
sampling time $t_F$.  Fig.~3 illustrates how we can find the horizontal 
angles $\phi$, having identified the rotation frequencies and angles $\theta$ for 
all control settings, using a local curve fitting approach similar to the second
strategy outlined in Sec.~\ref{sec:map} and described in Ref.~\cite{PRA69n050306}.
Finally, Fig.~4 shows how we can extract $\vec{d}_0$ and $\vec{d}_m$ by 
plotting the $x$, $y$ and $z$-components of the rotation axes $\vec{d}_0+f_m^{(k)}
\vec{d}_m$ versus $f_m^{(k)}$ for $m=1,2$ and fitting straight lines to the data.

\begin{figure}
\myincludegraphics[width=3.55in]{figures/pdf/Fig3.pdf}{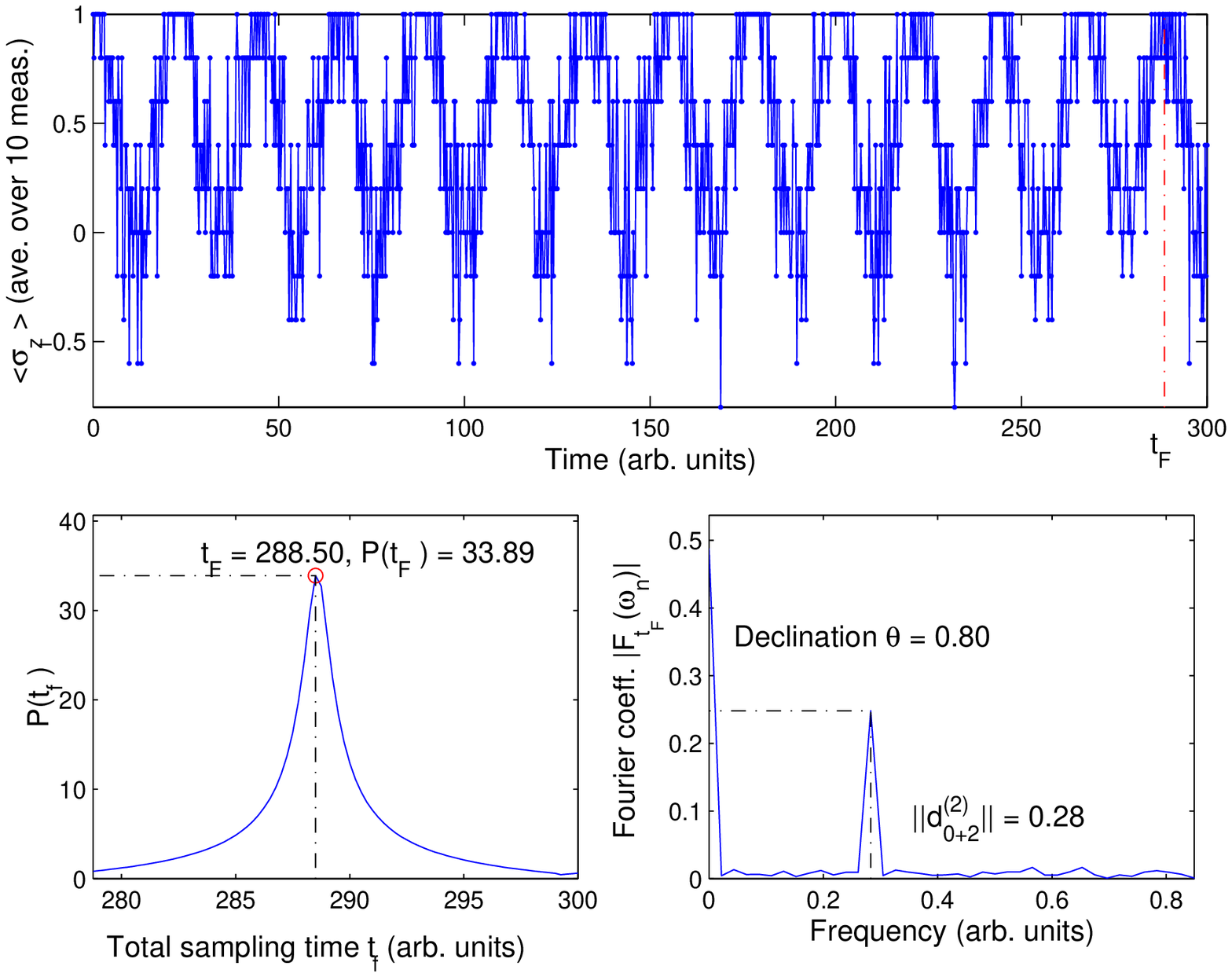}
\hspace{1em}\begin{minipage}[b]{2.6in}
{\bf Figure 2:}
The top graph to the left shows an example of noisy (simulated) measurement data
for rotations about $\vec{d}_0+f_2^{(2)}\vec{d}_2$ with $f_2^{(2)}=0.1$.  The  
dash-dot line indicates the optimal sampling time $t_F$ determined by finding the 
maximum of the function $P(t_f)$ (bottom-left).  The Fourier transform of the 
(truncated) data for $0\le t\le t_F$ is shown in the bottom-right graph.  Its 
zero-frequency component determines $\theta=\arccos\sqrt{|F(0)|}$; and the location
of the second peak gives the rotation frequency $\omega=\norm{\vec{d}_0+f_2^{(2)}
\vec{d}_2}$.  The estimates obtained, $\theta=0.80$ and $\omega=0.28$, are close
to the actual values $\theta=0.7854$ and $\omega=0.2828$.\\[2ex]
\end{minipage}
\end{figure}

\begin{figure}
\begin{minipage}[b]{3.2in}
{\bf Figure 3:}
The system is initialized in state $\vec{s}_1=(\cos\beta,\sin\beta,0)^T$ (here 
$\beta=-1.0446$) by rotating $\vec{s}_0=(0,0,1)^T$ about the reference axis 
$\vec{d}_r$ (here $\vec{d}_0$) by a suitable angle $\psi$, and the precession 
of $\vec{s}_1$ about $\vec{d}_0+f_1^{(1)}\vec{d}_1$, whose frequency and 
$\theta$ are already known, is mapped.  Using a small number of data points 
(stars) we find the $x$-intercept of $z(t)$ (circle), which allows us to 
estimate the location of the extrema of $z(t)$.  Additional data points (dots)
are then acquired in these regions, and parabolas fitted to the data.  The
desired angle $\phi=-\beta-\arcsin(\gamma \cos\delta/\sin\theta_r)$, where 
$\gamma=(z_{max}-z_{min})/2$ and $\delta=\pi-(\alpha_{min}+\alpha_{max})/2$ 
are determined by the vertices of the parabolas, and the values $\beta$ and 
$\theta_r$ (known from part 1).  We obtain $\phi=0.34$, which is close to the
actual value $\phi=0.3218$. 
\end{minipage}
\hspace{1em}
\myincludegraphics[width=3in]{figures/pdf/Fig4.pdf}{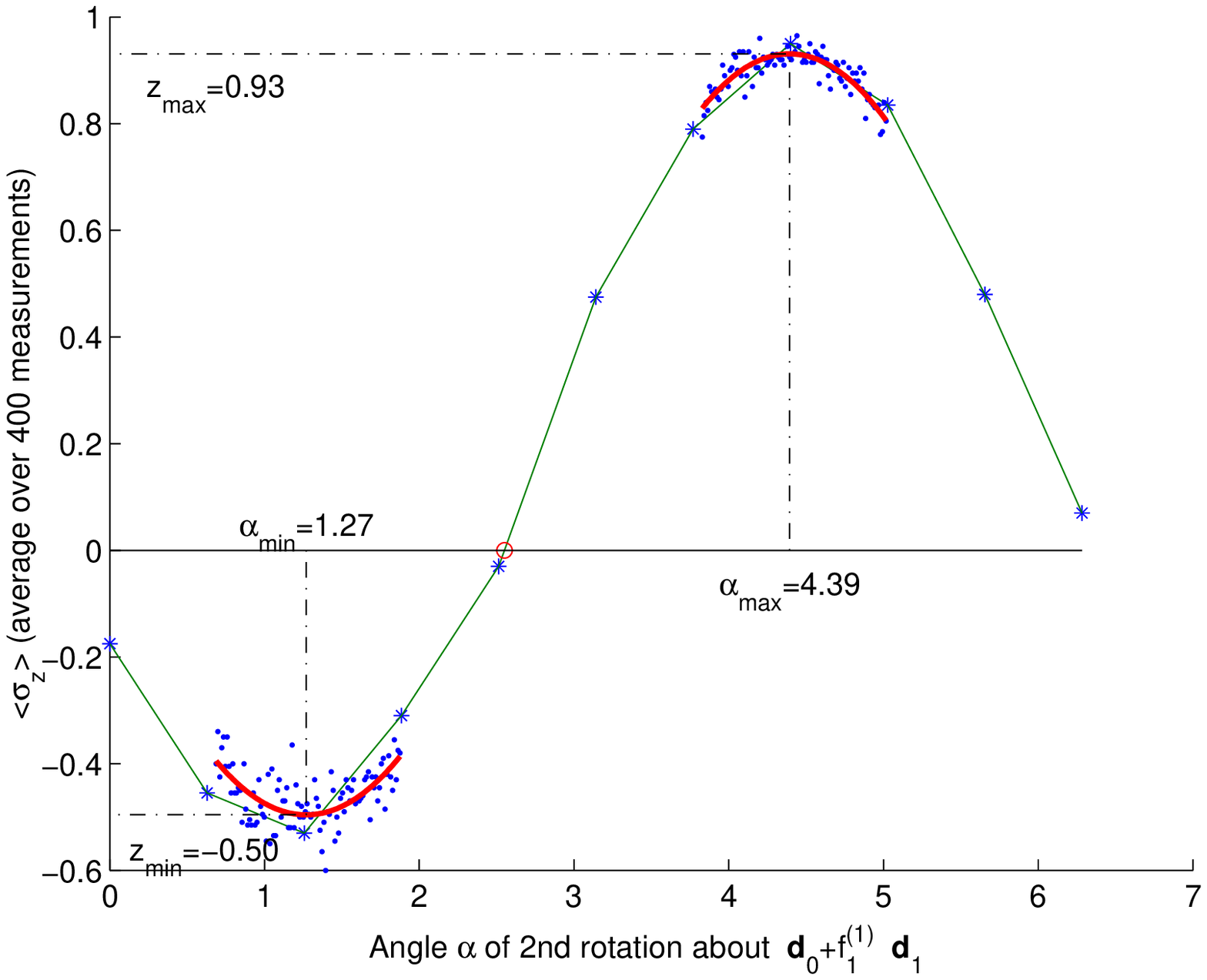}
\end{figure}

\begin{figure}
\myincludegraphics[width=3.6in]{figures/pdf/Fig5.pdf}{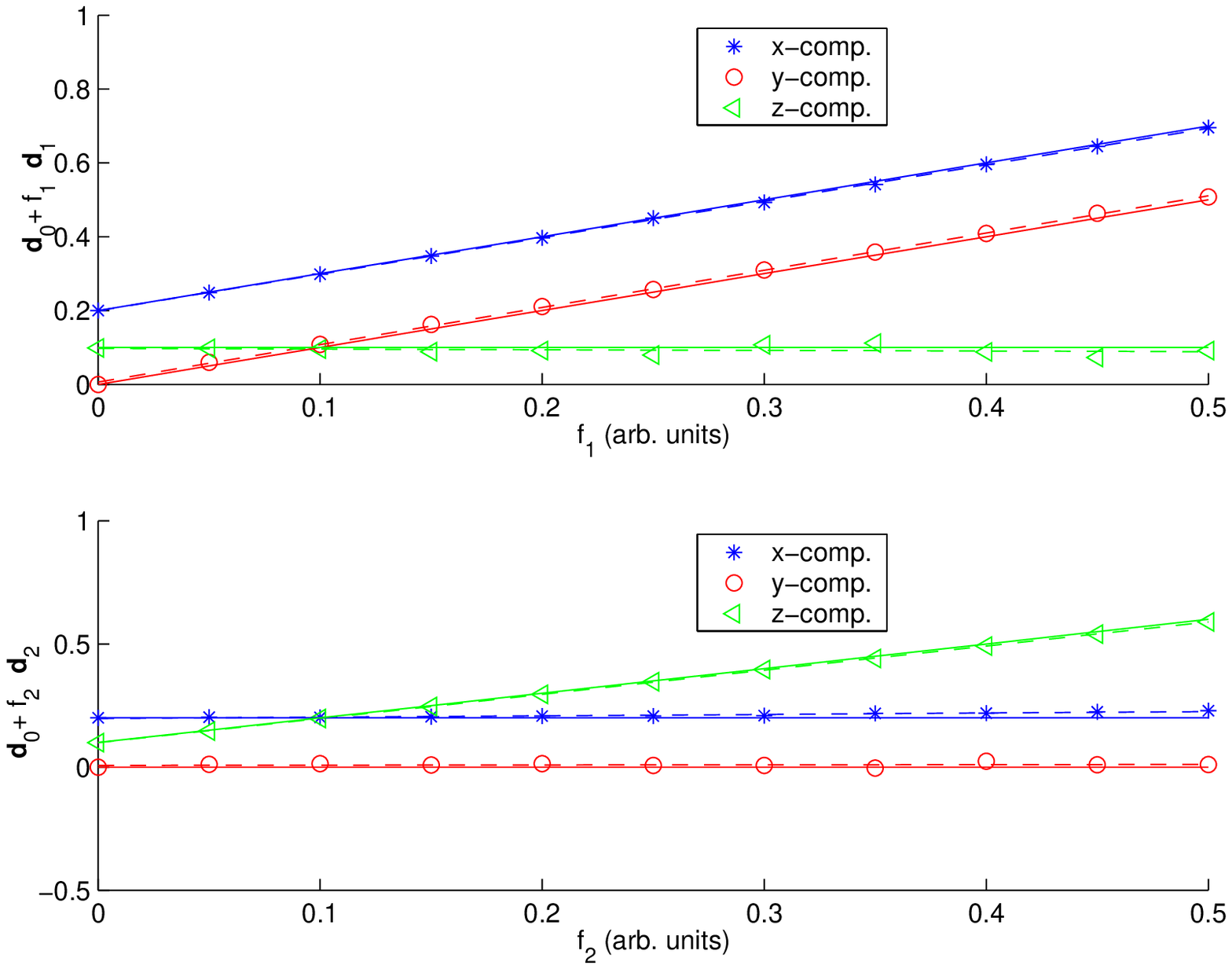}
\hspace{1em}\begin{minipage}[b]{2.85in}
{\bf Figure 4:}
Having determined the rotation frequencies $\omega_{0+m}^{(k)}$ and the angles
$\theta_{0+m}^{(k)}$ and $\phi_{0+m}^{(k)}$ of the rotation axes $\vec{d}_0+
f_m^{(k)}\vec{d}_m$ for various values of the controls $f_1$ and $f_2$, we 
convert the data into Cartesian coordinates, plot the values of the $x$, $y$
and $z$-components of the axes $\vec{d}_0+f_m^{(k)}\vec{d}_m$ for $m=1,2$, 
respectively, and fit straight lines.  The $y$-intercepts determine the $x$, 
$y$ and $z$-components of $\vec{d}_0$, the slopes those of $\vec{d}_m$, for 
$m=1,2$. For the data shown, we have
\begin{eqnarray*}
\vec{d}_0^{est}&=&(0.1986,0.0048,0.0979)^T \\
\vec{d}_1^{est}&=&(0.9884,1.0163,0.0087)^T \\
\vec{d}_2^{est}&=&(0.0531,0.0246,0.9819)^T
\end{eqnarray*}
The distances $\norm{\vec{d}_m^{est}-\vec{d}_m^{act}}$ of $0.0054$, $0.0218$ 
and $0.0613$, respectively, compare favorably to the 3\% readout error rate 
of the simulated experiments.
\end{minipage}
\end{figure}

%


\begin{theacknowledgments}
We thank A.~D. Greentree and L.~C.~H.~Hollenberg for helpful discussions.  S.G.S 
and D.K.L.O acknowledge financial support from the Cambridge-MIT Institute, Fujitsu, 
the UK goverment and IST grants RESQ (IST-2001-37559) and TOPQIP (IST-2001-39215).  
J.H.C acknowledges the support of the Australian Research Council, the Australian 
government, the US National Security Agency, The Advanced Research and Development
Activity and the US Army Research Office (DAD19-01-1-0653).  D.K.L.O also thanks 
Sidney Sussex College for support.  
\end{theacknowledgments}


\bibliographystyle{aipproc}   

\bibliography{papers,sonia}

\IfFileExists{\jobname.bbl}{}
 {\typeout{}
  \typeout{******************************************}
  \typeout{** Please run "bibtex \jobname" to optain}
  \typeout{** the bibliography and then re-run LaTeX}
  \typeout{** twice to fix the references!}
  \typeout{******************************************}
  \typeout{}
 }

\end{document}